\documentclass[12pt]{article}
\usepackage{amssymb}
\usepackage{latexsym}
\usepackage{amsmath}
\usepackage{graphicx}

\def\vp{\varphi}
\def\th{\theta}
\def\Dl{\Delta}
\title{\LARGE{Charged fermion tunnelling  from electrically and magnetically charged rotating black hole in de Sitter space}}
\author{M. M. Stetsko\footnote{E-mail: mstetsko@gmail.com, mykola@ktf.franko.lviv.ua}\
\\
  {\small Department of Theoretical Physics, Ivan Franko National University of Lviv,}\\
{\small 12 Drahomanov Str., Lviv, UA-79005, Ukraine
         }}

\begin{document}
\maketitle

\abstract{Thermal radiation of electrically charged fermions from
rotating black hole with electric and magnetic charges in de
Sitter space is considered. The tunnelling probabilities for
outgoing and incoming particles are obtained and the Hawking
temperature is calculated. The relation for the classical action
for the particles in the black hole's background is also found.}

\section{Introduction}
Hawking radiation has been attracting a lot of attention since it
was proposed \cite{Hawking_CMP75}. To consider it different
methods have been applied \cite{Brout_PRept95}. The semi-classical
tunnelling approach that was proposed by Kraus and Wilczek
\cite{Kraus_Wilczek,Kraus_Wilczek_2} has gained considerable
interest recently. It was shown that Hawking temperature is
defined by the imaginary part of the emitted particle's action for
the classically forbidden region near the horizon. To calculate it
two methods were proposed. The first one is so-called null
geodesic method proposed by Parikh and Wilczek \cite{Parikh_PRL00}
is based on the fact that imaginary part of the action  is caused
by the integration of radial momentum $p_r$  for the emitted
particles. The second method is based on relativistic
Hamilton-Jacobi equation, and the imaginary part of the action can
be obtained after integration of that equation
\cite{Angheben_JHEP05}. This second approach can be treated as an
extension of complex path method proposed by Padmanabhan with
collaborators \cite{Padmanabhan_PRD99}-\cite{Padmanabhan_CQG02}

Tunnelling approach based on Hamilton-Jacobi equation at first was
applied to the emission of scalar particles. Then it was
successfully applied to vast area of well-known and exotic
space-times, in particular Kerr and Kerr-Newman ones
\cite{Q_Jiang_PRD06, J_Zhang_PLB06}, Taub-NUT space-time
\cite{kerner_PRD07} G\"{o}del space-time \cite{kerner_PRD07_2},
BTZ black holes \cite{S_Wu_JHEP06}, dynamical black holes
\cite{di_Criscienzo_PLB07}. The review of tunnelling method is
considered in paper \cite{Vanzo_CQG11} where further references
can be found.

Tunnelling approach was also successfully applied to tunnelling of
fermions. In their seminal work Kerner and Mann used Dirac
equation instead of Hamiton-Jacobi one  to obtain temperature of
emitted fermions and showed that for a chosen type of space-time
temperature of emitted fermions would be the same as the
temperature of scalar particles \cite{Kerner_CQG08}. That method
was later applied to different kinds of black hole space-times
including Reissner-Nordstr\"{o}m one \cite{X_Zeng_GRG08},
Kerr-Newman \cite{Chen_PLB08, Kerner_PLB08}, dilatonic black holes
\cite{Chen_CQG08}, BTZ black hole \cite{Li_PLB08}, black holes in
Ho\v{r}ava-Lifshitz gravity \cite{Chen_PLB09, Liu_CQG11},
accelerating and rotating black hole \cite{Rehnam_JCAP11,
Sharif_EPJC12}, rotating black strings \cite{Ahmed_JCAP11}.

In our work we consider Kerr-Newman-de Sitter black  hole which
carries both electric and magnetic charges. Using Kerner-Mann
procedure we consider emission of charged spin $1/2$ particles. We
show that in the presence of electric and magnetic charges of the
black hole the variables of Dirac equation can also be separated
and as a consequence the temperature can be found. We also find
the quasiclassical action of emitted particle.

\section{Charged spin $1/2$ particle tunnelling from Kerr-Newman-de Sitter black hole}
Emission of charged particles with spin $1/2$ was considered
independently in works \cite{Chen_PLB08, Kerner_PLB08}. We
consider emission of charged spin $1/2$ from Kerr-Newman-de Sitter
black hole which carries both electric and magnetic charges
(dyonic black hole). Kerr-Newman-de Sitter black hole's metrics in
Boyer-Lindquist coordinates takes form:
\begin{equation}
ds^2=-\frac{\Dl}{\rho^2}\left(dt-\frac{a\sin^2{\th}}{\Xi}d\vp\right)^2+\frac{\rho^2}{\Dl}dr^2
+\frac{\rho^2}{\Dl_{\th}}d\th^2+\frac{\Dl_{\th}\sin^2{\th}}{\rho^2}\left(\frac{R^2}{\Xi}d\vp-adt\right)^2,
\end{equation}
where $R^2=r^2+a^2$, $\rho^2=r^2+a^2\cos^2{\th}$,
$\Dl=R^2(1-r^2/l^2)+Q_e^2+Q_m^2-2Mr$, $\Xi=1+a^2/l^2$,
$\Dl_{\th}=1+a^2/l^2\cos^2{\th}$. Here $a=J/M$ is the angular
momentum parameter and another parameter $l^2=3/\Lambda$ is
defined by the cosmological constant, parameters $Q_e$ and $Q_m$
are electric and magnetic charges respectively. It is known that
black hole's horizons can be found from equation $\Dl=0$ and in
the case of Kerr-Newman-de Sitter metrics we have four roots. The
largest one is the cosmological horizon $r_c$ (CH), the minimal
positive root is the Cauchy horizon $r_i$ and the intermediate
root is the event horizon $r_+$. The last root is negative and it
is not taken into consideration.

Components of electromagnetic potential takes the form:
\begin{eqnarray}
A_t=-\frac{1}{\rho^2}(Q_er+Q_ma\cos{\th}),
\\
A_{\vp}=\frac{1}{\Xi\rho^2}(Q_era\sin^2{\th}+Q_mR^2\cos{\th})
\end{eqnarray}

The Dirac equation for electrically charged particle takes form:
\begin{equation}\label{Dirac_General}
i\gamma^{\mu}\left(D_{\mu}-\frac{iq}{\hbar}A_{\mu}\right)\psi+\frac{m}{\hbar}\psi=0
\end{equation}
where $D_{\mu}=\partial_{\mu}+\Omega_{\mu}$,
$\Omega_{\mu}=\frac{1}{8}\Gamma^{\alpha\beta}_{\mu}[\gamma^{\beta},\gamma^{\alpha}]$
and $\gamma^{\mu}$ matrices are obeyed to commutation relation:
\begin{equation}
[\gamma^{\mu},\gamma^{\nu}]=2g^{\mu\nu}\hat{1}
\end{equation}

The representation for the $\gamma^{\mu}$ can be taken as follows:

\begin{eqnarray}\label{gamma_repr}
\gamma^t=\frac{\sqrt{K(r,\th)}}{\rho\sqrt{\Dl\Dl_{\th}}}\gamma^0,
\quad
\gamma^r=\frac{\sqrt{\Dl}}{\rho}\gamma^3, \quad \gamma^{\th}=\frac{\sqrt{\Dl_{\th}}}{\rho}\gamma^1,\\
\gamma^{\vp}=\frac{\Xi}{\sqrt{K(r,\th)}}\left(\frac{\rho}{\sin{\th}}
\gamma^2-\frac{a(\Dl-R^2\Dl_{\th})}{\rho\sqrt{\Dl\Dl_{\th}}}\gamma^0\right)
\end{eqnarray}
where we denoted $K(r,\th)=R^4\Dl_{\th}-\Dl a^2\sin^2{\th}$ and
matrices $\gamma^a$ take following form:
\begin{eqnarray}
\gamma^0=
\begin{pmatrix}
0 & I \\
-I & 0\\
\end{pmatrix} \quad
\gamma^1=
\begin{pmatrix}
0 & \sigma_1 \\
\sigma_1 & 0\\
\end{pmatrix}\\
\gamma^2=
\begin{pmatrix}
0 & \sigma_2 \\
\sigma_2 & 0\\
\end{pmatrix} \quad
\gamma^3=
\begin{pmatrix}
0 & \sigma_3 \\
\sigma_3 & 0\\
\end{pmatrix}
\end{eqnarray}
and $\sigma_i$ are the Pauli matrices. We note that representation
similar to (\ref{gamma_repr})was used in the work
\cite{Kerner_PLB08}.

We choose the wave functions with spin up and down in the form:
\begin{eqnarray}\label{wave_func_up}
\psi_{\uparrow}=
\begin{pmatrix}
A(t,r,\th,\vp)\\ 0 \\ B(t,r,\th,\vp)\\0\\
\end{pmatrix}
\exp{\left(\frac{i}{\hbar}I_{\uparrow}(t,r,\th,\vp)\right)}\quad
\psi_{\downarrow}=
\begin{pmatrix}
0 \\C(t,r,\th,\vp) \\0 \\B(t,r,\th,\vp)\\
\end{pmatrix}
\exp{\left(\frac{i}{\hbar}I_{\downarrow}(t,r,\th,\vp)\right)},
\end{eqnarray}
where $I_{\uparrow}$ and $I_{\downarrow}$ are the action for the
Dirac particles with spin-up ($\uparrow$) and spin-down
($\downarrow$) tunnelling through the horizons.

Then we substitute (\ref{wave_func_up}) into Dirac equation
(\ref{Dirac_General}) and performing quasiclassical approximation
we obtain
\begin{eqnarray}\label{dirac_1}
B\left[-\frac{\sqrt{K(r,\th)}}{\rho\sqrt{\Dl\Dl_{\th}}}\partial_tI_{\uparrow}-\frac{\sqrt{\Dl}}{\rho}\partial_r
I_{\uparrow}+\frac{\Xi a(\Dl-R^2\Dl_{\th})}{\rho\sqrt{\Dl\Dl_{\th}
K(r,\th)}}\partial_{\vp}I_{\uparrow}-\frac{q\sqrt{(K(r,\th))}}{\rho^3\sqrt{\Dl\Dl_{\th}}}
(Q_er+Q_ma\cos{\th})\right.\\\nonumber
\left.-\frac{qa(\Dl-R^2\Dl_{\th})}{\Xi\rho^3\sqrt{\Dl\Dl_{\th}
K(r,\th)}}(Q_era\sin^2{\th}+Q_mR^2\cos{\th})\right]+mA=0;
\end{eqnarray}
\begin{eqnarray}\label{dirac_2}
B\left[-\frac{\sqrt{\Dl_{\th}}}{\rho}\partial_{\th}I_{\uparrow}-\frac{i\Xi\rho}{\sqrt{K(r,\th)}\sin{\th}}
\partial_{\vp}I_{\uparrow}+\frac{iq}{\Xi\rho\sqrt{K(r,\th)}\sin{\th}}
(Q_era\sin^2{\th}+Q_mR^2\cos{\th})\right]=0;
\end{eqnarray}
\begin{eqnarray}\label{dirac_3}
A\left[\frac{\sqrt{K(r,\th)}}{\rho\sqrt{\Dl\Dl_{\th}}}\partial_tI_{\uparrow}-\frac{\sqrt{\Dl}}{\rho}\partial_r
I_{\uparrow}-\frac{\Xi a(\Dl-R^2\Dl_{\th})}{\rho\sqrt{\Dl\Dl_{\th}
K(r,\th)}}\partial_{\vp}I_{\uparrow}+\frac{q\sqrt{(K(r,\th))}}{\rho^3\sqrt{\Dl\Dl_{\th}}}
(Q_er+Q_ma\cos{\th})\right.\\\nonumber
\left.+\frac{qa(\Dl-R^2\Dl_{\th})}{\Xi\rho^3\sqrt{\Dl\Dl_{\th}
K(r,\th)}}(Q_era\sin^2{\th}+Q_mR^2\cos{\th})\right]+mB=0;
\end{eqnarray}
\begin{eqnarray}\label{dirac_4}
B\left[-\frac{\sqrt{\Dl_{\th}}}{\rho}\partial_{\th}I_{\uparrow}-\frac{i\Xi\rho}{\sqrt{K(r,\th)}\sin{\th}}
\partial_{\vp}I_{\uparrow}+\frac{iq}{\Xi\rho\sqrt{K(r,\th)}\sin{\th}}
(Q_era\sin^2{\th}+Q_mR^2\cos{\th})\right]=0.
\end{eqnarray}
Note, that in the first order WKB approximation the terms
proportional to $\Omega_{\mu}$ are omitted.

Suppose that the action $I_{\uparrow}$ takes the form:
\begin{equation}\label{action}
I_{\uparrow}=-Et+J\vp+W(r,\th)
\end{equation}
Now we substitute (\ref{action}) into
(\ref{dirac_1})-(\ref{dirac_4}). In order to make these equations
simpler they are decomposed into the series near the horizon
surface \cite{Chen_PLB08, Kerner_PLB08} (here the decomposition in
the vicinity of the event horizon is represented):
\begin{eqnarray}\label{dirac_1h}
B\left[\frac{R_{+}^2E-\Xi
aJ-qQ_er_+}{\rho_{+}\sqrt{\Dl_r(r_+)(r-r_+)}}-\frac{\sqrt{\Dl_r(r_+)(r-r_+)}}{\rho_+}W_r
(r,\th)\right]+mA=0;
\end{eqnarray}
\begin{eqnarray}\label{dirac_2h}
B\left[-\frac{\sqrt{\Dl_{\th}}}{\rho_+}W_{\th}(r,\th)-\frac{i}{R_+^2\sqrt{\Dl_{\th}}}\left(\frac{\Xi\rho_+J}{\sin{\th}}-\frac{q}{\Xi\rho_+}
(Q_er_+a\sin^2{\th}+Q_mR_+^2\cos{\th})\right)\right]=0
\end{eqnarray}
\begin{eqnarray}\label{dirac_3h}
A\left[-\frac{R_{+}^2E-\Xi
aJ-qQ_er_+}{\rho_{+}\sqrt{\Dl_r(r_+)(r-r_+)}}-\frac{\sqrt{\Dl_r(r_+)(r-r_+)}}{\rho_+}W_r
(r,\th)\right]+mB=0;
\end{eqnarray}
\begin{eqnarray}\label{dirac_4h}
A\left[-\frac{\sqrt{\Dl_{\th}}}{\rho_+}W_{\th}(r,\th)-\frac{i}{R_+^2\sqrt{\Dl_{\th}}}\left(\frac{\Xi\rho_+J}{\sin{\th}}-\frac{q}{\Xi\rho_+}
(Q_er_+a\sin^2{\th}+Q_mR_+^2\cos{\th})\right)\right]=0.
\end{eqnarray}
Here
$\Dl_r(r_+)=2\left(r_+\left(1-a^2/l^2\right)-2r_+^3/l^2-M\right)$,
$R_+^2=r_+^2+a^2$ and $\rho_+^2=r_+^2+a^2\cos^2{\th}$.

For the massless case equations (\ref{dirac_1h}) and
(\ref{dirac_3h}) decouple and can be solved. It is easy to see
that variables $r$ and $\th$ can be separated. So the function
$W(r,\th)$ is represented in form:
\begin{equation}\label{separ_var}
W(r,\th)=W(r)+\Theta (\th)
\end{equation}
When $A=0$ equation (\ref{dirac_1h}) leads to:
\begin{equation}
W_r(r,\th)=\frac{R_+^2E-\Xi aJ-qQ_er_+}{\Dl_r(r_+)(r-r_+)}
\end{equation}
Having integrated around the pole and taking the imaginary part of
the action we obtain:
\begin{equation}\label{imag_part_W+}
Im W_+=\frac{\pi
R_+^2\left(E-\Xi\Omega_+J-qQ_e\frac{r_+}{R_+^2}\right)}{\Dl_r(r_+)},
\end{equation}
where $\Omega_+=a/R_+^2$ is the angular velocity at the horizon.

Similarly when $B=0$ we write:
\begin{equation}
W_r(r,\th)=-\frac{R_+^2E-\Xi aJ-qQ_er_+}{\Dl_r(r_+)(r-r_+)}
\end{equation}
And after integration the result is as follows:
\begin{equation}
Im W_-=-\frac{\pi
R_+^2\left(E-\Xi\Omega_+J-qQ_e\frac{r_+}{R_+^2}\right)}{\Dl_r(r_+)}
\end{equation}
As it was argued in \cite{Padmanabhan_PRD99, Padmanabhan_MPLA01,
Padmanabhan_CQG02, Kerner_PLB08} that probabilities of crossing
the horizon are defined by the imaginary part of the action:
\begin{equation}\label{prob}
P_{out}\propto\exp[-2(ImW_++Im\Theta)], \quad
P_{in}\propto\exp[-2(ImW_-+Im\Theta)].
\end{equation}
The resulting tunnelling probability is represented as the ratio
of probabilities (\ref{prob}):
\begin{equation}\label{gamma_def}
\Gamma\propto\frac{P_{out}}{P_{in}}=\frac{\exp[-2Im
W_+]}{\exp[-2Im W_-]}=\exp[-4ImW_+].
\end{equation}
Substituting relation (\ref{imag_part_W+}) into (\ref{gamma_def})
we obtain:
\begin{equation}
\Gamma=\exp\left(-4\pi\frac{R_+^2\left(E-\Xi\Omega_+
J-qQ_e\frac{r_+}{R_+^2}\right)}{\Dl_r(r_+)}\right).
\end{equation}
As a result the temperature takes form:
\begin{equation}\label{temperature}
T=\frac{\Dl_r(r_+)}{4\pi
R_+^2}=\frac{\left(r_+\left(1-a^2/l^2\right)-2r_+^3/l^2-M\right)}{2\pi(r_+^2+a^2)}.
\end{equation}
 In the massive case ($m\neq
0$) one has to solve the system of equations (\ref{dirac_1h}) and
(\ref{dirac_3h}). So we obtain:
\begin{equation}
\frac{2AB}{\rho_+^2}\left[\frac{R_+^2E-\Xi
aJ-qQ_er_+}{\sqrt{\Dl_r(r_+)(r-r_+)}}\right]+m(A^2-B^2)=0
\end{equation}
and as a result:
\begin{equation}
\frac{A}{B}=\frac{-(R_+^2E-\Xi aJ-qQ_er_+)\pm\sqrt{(R_+^2E-\Xi
aJ-qQ_er_+)^2+m^2\rho_+^2\Dl_r(r_+)(r-r_+)}}{m\rho_+\sqrt{\Dl_r(r_+)(r-r_+)}}
\end{equation}
When $r\rightarrow r_+$ ratio $A/B$ can tend to $0$ or $-\infty$
(see also \cite{Kerner_PLB08}). When $A/B\rightarrow 0$ then
$A\rightarrow 0$ the equation (\ref{dirac_3h}) is solved in terms
of $m$ and the result is inserted into (\ref{dirac_1h}). It is
easy to see that resulting expression does not depend on the
variable $\th$. So we obtain:
\begin{equation}\label{radial_der_massive_case}
W_r(r,\th)=W_+'(r)=\frac{R_+^2\left(E-\Xi
\Omega_+J-qQ_e\frac{r_+}{R_+^2}\right)}{\Dl_r(r_+)(r-r_+)}\frac{1+\frac{A^2}{B^2}}{1-\frac{A^2}{B^2}}
\end{equation}
The result that is found after integration around the pole is the
same as in the massless case since $A/B\rightarrow 0$ at the
horizon. When $B\rightarrow 0$ the relation
(\ref{radial_der_massive_case}) can be rewritten as follows:
\begin{equation}\label{radial_der_massive_case_2}
W_r(r,\th)=W_-'(r)=-\frac{R_+^2\left(E-\Xi
\Omega_+J-qQ_e\frac{r_+}{R_+^2}\right)}{\Dl_r(r_+)(r-r_+)}\frac{1+\frac{B^2}{A^2}}{1-\frac{B^2}{A^2}}
\end{equation}
Similarly the result after integration gets no correction in
addition to the massless case. So as a consequence the temperature
for emitted massive fermions is the same as for the massless and
given by the relation (\ref{temperature}).

It is known that Hawking radiation in de Sitter space can also
appear at the cosmological horizon. In contrast to the emission of
particles at the event horizon Hawking radiation at the
cosmological horizon is caused by incoming particles whereas
outgoing particles move along classically permitted trajectories.
In spite of that qualitative difference calculations of the
Hawking temperature at the cosmological horizon can be made in the
same way as at the event horizon. To obtain the expression for the
Hawking temperature at the cosmological horizon one should replace
the event horizon radius in formula (\ref{temperature}) by the
cosmological one ($r_+\rightarrow r_c$). So we have:
\begin{equation}\label{temperature_cosmol}
T=\frac{\Dl_r(r_c)}{4\pi
R_c^2}=\frac{\left(r_c\left(1-a^2/l^2\right)-2r_c^3/l^2-M\right)}{2\pi(r_c^2+a^2)}.
\end{equation}
Relations for temperature at the event horizon (\ref{temperature})
as well as at the cosmological one take the same form as in case
of the hole with only the electric charge \cite{Chen_PLB08}. It
should be noted that radii of horizons depend on the both electric
and magnetic charges so our formulas are consistent with relations
given in \cite{Chen_PLB08} when $Q_m\rightarrow 0$.

\section{Action for the emitted particles}
Equations (\ref{dirac_1h})-(\ref{dirac_4h}) allow one to obtain
explicit expression for action of emitted particles. As it was
already shown angle and radial variables are separated near the
horizon. So the radial and angular parts of the action can be
obtained independently. Having used the relation (\ref{separ_var})
the equation (\ref{dirac_1h}) can be rewritten in the form:
\begin{equation}
W'(r)=\frac{R_{+}^2E-\Xi
aJ-qQ_er_+}{\Dl_r(r_+)(r-r_+)}+\frac{Am\rho_+}{B\sqrt{\Dl_r(r_+)(r-r_+)}}
\end{equation}
After integration we obtain:
\begin{equation}\label{action_radial_part_+}
W_+(r)=\frac{R_{+}^2E-\Xi
aJ-qQ_er_+}{\Dl_r(r_+)}\ln(r-r_+)+\int\frac{Am\rho_+}{B\sqrt{\Dl_r(r_+)(r-r_+)}}dr.
\end{equation}
In order to find relation for radial part of the action for
incoming particles one should integrate equation (\ref{dirac_3h}).
After integration we arrive at the following relation:
\begin{equation}\label{action_radial_part_-}
W_-(r)=-\frac{R_{+}^2E-\Xi
aJ-qQ_er_+}{\Dl_r(r_+)}\ln(r-r_+)+\int\frac{Bm\rho_+}{A\sqrt{\Dl_r(r_+)(r-r_+)}}dr.
\end{equation}
For the angular part relation (\ref{dirac_2h}) can be represented
in the form:
\begin{equation}
\Theta'(\th)=-\frac{i\Xi\rho_+^2J}{R_+^2\sin{\th}\Dl_{\th}}+\frac{iq(Q_er_+a\sin^2{\th}+Q_mR_+^2\cos{\th})}{\Xi
R_+^2\Dl_{\th}}
\end{equation}
Having integrated the last equation we obtain:
\begin{eqnarray}\label{action_angular_part}
\Theta(\th)=\frac{iqQ_ml^2}{\Xi
a\sqrt{l^2-a^2}}\arctan{\left[\frac{l\sin{\th}}{\sqrt{l^2-a^2}}\right]}+\frac{i\Xi
Jal}{2(a^2-l^2)}\left(\ln\left|\frac{l+a\cos{\th}}{l-a\cos{\th}}\right|-
\frac{l}{a}\ln\left|\frac{1+\cos{\th}}{1-\cos{\th}}\right|\right)\nonumber
\\
-\frac{i\Xi
Jal}{2R_+^2}\ln\left|\frac{l+a\cos{\th}}{l-a\cos{\th}}\right|+\frac{iqQ_ear_+}{\Xi
R_+^2}\left(\frac{l^2}{a^2}\th-\frac{1}{a}\sqrt{\l^2-a^2}\arctan\left[\frac{l\tan{\th}}{\sqrt{l^2-a^2}}\right]\right)
\end{eqnarray}
Using relations (\ref{action_radial_part_+}) and
(\ref{action_angular_part}) one can write the action for outgoing
massive particles. Similarly equations
(\ref{action_radial_part_-}) and (\ref{action_angular_part}) allow
one to get the action for incoming particles.

\section{Conclusions}
In  this paper we considered charged fermion tunnelling form the
electrically and magnetically charged Kerr-Newman-de Sitter black
hole. Using Kerner-Mann approach \cite{Kerner_PLB08} we
successfully recovered black hole's temperature. It was shown that
similarly to the case when black carries only the electric charge
inclusion additional magnetic charge does not spoil separability
of the Dirac equation in the vicinity of the horizons. So
relations for the temperature is obtained in the same manner and
take almost the same form as it was in the case of electrically
charged black hole. We also note that for temperature explicit
dependence on the magnetic charge is hidden in definition of
horizons radii.

We also obtained relations for radial and angular part of the
action. Those relations might be helpful if one tries to find
corrections to the spectrum of emitted particles. Here we have
explicit dependence on electric as well as magnetic charges so
these terms might have different influence on the spectrum of
emitted particles.

Another issue that still remains open is taking into account
higher order of WKB corrections. This problem is connected with
calculation of terms caused by spin connection. These terms can
affect on the separability and tractability of the Dirac equation
and this problem requires additional careful consideration.

\end{document}